\documentstyle[12pt]{article}
\title{Black Holes in Brane Worlds}
\author{P. Suranyi and L.C.R. Wijewardhana\\Department of Physics,
University of Cincinnati, Cincinnati, Ohio, 45221}

\begin{document}
\maketitle

\begin{abstract}
   In a Randall-Sundrum theory (RS1)
3+1 dimensional black holes and higher dimensional black holes are not the
natural continuations of each other.  3+1 dimensional black holes decay
into a large number of 4+1 dimensional black holes at a critical
mass, $M_{\rm crit}\sim 10^{32}$ TeV.  Those black holes
themselves may become unstable above another, albeit much smaller critical
mass, $M_0\sim 10^3$TeV.

\end{abstract}
\maketitle
 
Models of the universe with extra dimensions larger than the Planck length
have been under intense investigation during the last few years
\cite{add1, rs1, rs2, rubakov}. The general feature of these models is
that standard model particles are compelled to live on 3-branes, to satisfy
momentum conservation in 3+1 dimensions and to conform to other
phenomenological bounds, while gravity pervades all dimensions.
     Many of these models predict the observation of black holes
  at future accelerators~\cite{add1,giddings,argyres,shapere}.  The
models either use flat but compact extra dimensions(ADD scenario)~\cite{add1} or
   a number of branes embedded in AdS space, with  warped extra
dimension(s)~\cite{rs1,rs2}.

Black holes in theories with extra dimensions have
   been studied
widely.  The classic paper of Myers and Perry~\cite{myers} found
solutions in  $D$-dimensional flat space.  Black hole solutions were also
found
in AdS space~\cite{hawking1,birmingham}. No non-trivial  black hole
solutions
have been found in closed form in brane theories of the Randall Sundrum
type. Yet, it is
important to learn as much as possible about  black holes in such models.
   The black string solution~\cite{hawking2} that  extends in a uniform
manner from the brane into the extra dimension  has the Gregory-Laflamme
instability in the ADD scenario~\cite{hawking2,gregory1}.  It is easy to invoke an entropy
argument  ~\cite{gregory2}, to show that an instability will occur
at a critical mass. An alternative interpretation is given
in~\cite{horowitz}.

   To understand the arguments by Gregory and Laflamme, compare the
entropies of standard 3+1   and 4+1
dimensional Schwarzschild black holes of the same mass. Then one obtains a critical mass
\[
M_{\rm crit}\sim \frac{M_4^4}{M_5^3},
\]
where $M_D$is the Planck mass in $D$ dimensions.
At this mass the radius of the horizon of the 4 dimensional black hole is
approximately the same as that of the   5 dimensional one. When a 4 dimensional
black hole in the process of its Hawking radiation~\cite{hawking3,x}
passes this critical mass,  the entropy of a 5
dimensional black hole with the same mass becomes larger.
Then, according to Gregory and Laflamme, under the influence of quantum
fluctuations, the black string breaks up into a large number of small 5
dimensional black holes which then unite into a single 5 dimensional
black hole, having a larger entropy than the collection of smaller black
holes.

In what follows we would like to investigate the question of stability of
black holes in models with warped extra dimensions. We will consider both
of the models of Randall and Sundrum (RS1 and
RS2)~\cite{rs1,rs2}. We will use an expansion technique to find such black
hole solutions
and then discuss their properties.

  First we will investigate black holes in
RS1 and find a markedly different behavior from ADD.
One important goal of RS1 is to solve the hierarchy problem by making the fundamental gravity scale 1 TeV. This is
achieved by requiring an exponential relation between the scales on the TeV brane,
the home of standard model particles, and on the Planck brane.

We start form the Randall-Sundrum classical  action  $ S $   for the
system consisting of two 3-branes with
brane tensions $\sigma_{Planck} =  -  \sigma_{sm} ={12M^{3} \over{l} } $ .
The branes are fixed at the points $ y = 0 $  and $ y =y_{\rm max}=
\pi r_{c} $.

\begin{eqnarray}
S &=& \frac{M_5^{3}}{16\pi}\int d ^{4}x\int dy \sqrt{-g}\,( {12
\over{\lambda^{2}}} + R )\nonumber\\&+&
\int d^{4}x \sqrt{-\eta}\,\sigma_{sm} +
\int d^{4}x \sqrt{-\eta}\,\sigma_{Planck}
\label{action1}
\end{eqnarray}
where R is the 5-d Ricci scalar, $M_5$ is the fundamental gravity scale,
$\lambda$ is the curvature length of the AdS space  and
the brane tensions
$\sigma_{\rm TeV}$ and
$\sigma_{\rm Planck} $ are selected such that the metric satisfies the
correct induced Einstein equations on the branes.

    In RS1 the range of the 
variable
$y$ of (\ref{action1}) is limited to
$0\le y\le y_{\rm max}$. As we are interested in objects living on
the TeV brane, located at
$y=y_{\rm max}$ rather than on the Planck brane located at $y=0$ it is
convenient to introduce the conformal
variable
$w$~\cite{shapere} defined by $|w|=z_c-\lambda\exp\{y/\lambda\}$
where
$z_c=\lambda\exp\{y_{\rm max}/\lambda\}$. Note that unlike in
Ref.\cite{shapere} we use the convention $G=M_d^{-d+2}$. The range of
variable
$w$ is
$0<|w|<z_c-\lambda$. The location of the TeV brane, where standard model
particles live, is $w=0$. The parameter
$z_c=O(TeV^{-1})$.

    The metric in this coordinate system takes the form
\begin{equation}
ds^2=\left(\frac{z_c}{z_c-|w|}\right)^2\left(\eta_{\mu\nu}{dx'}^\mu{dx'}^\nu-dw^
2,\right)
\label{metric2}
\end{equation}
where the rescaled `brane variables' are defined as ${dx'}^\mu=\lambda
dx^\mu/z_c$. The new  5-d gravity action is given by

\begin{eqnarray}
S &=& \frac{\tilde M_{5}^{3}}{16\pi}\int d ^{4}x'\int dw \sqrt{-G}\left(
\frac{12}{ z_{c}^{2}} + R_{G}\right)+S_{\rm brane},
\label{action2}
\end{eqnarray}
where $ \tilde M_{5} $ =$ M_5{ \lambda \over z_c} $ and $R_G=\lambda^2
R/z_c^2$. The crucial difference between the actions  (\ref{action1})
and (\ref{action2}) is that the gravitational constant,
$M_5^3$ is rescaled to $\tilde M_5^3= (\lambda M_5/z_c)^3$, where $\tilde
M_5=O($1TeV) can be chosen and the new effective AdS curvature length is
$z_c$.  As we will deal with this rescaled 5 dimensional Planck mass only,
we will use the notation
$M_5$ for  $\tilde M_5$.

Let us now consider a black hole in the RS1 scenario bound to the TeV brane
with its singularity on the brane. As long as the
radius R of the horizon of such a black hole   is
substantially smaller than
$z_c$, the effect of the AdS term of (\ref{action2}) is
negligible. Thus, such an object could very well be described
by a flat 5 dimensional black hole solution. Then the constraint $R<<z_c$
can be
translated to the relation
\[
M<<M_0=M_53\pi(z_cM_5)^2/8.
\]
Using phenomenological constraints, the dimensionless constant $z_cM_5$
was estimated~\cite{shapere} to be $z_cM_5\ge 20$.  Setting the Planck
mass at 1 TeV this bound for the black hole mass is $M<<500$TeV. As
quantum gravity sets in at around 1 TeV black holes satisfying this bound
do exist and can presumably be produced at future
accelerators~\cite{shapere}.

Consider now a primordial RS1 black hole of mass 1-10TeV,  produced when
the temperature of the universe was $T\sim 1$TeV~\cite{note}.  The Hawking
temperature of the black hole,
\[
T_{\rm BH}=\sqrt{3M_5^3/32\pi M_{\rm
BH}},
\]
   for D = 5,
   will satisfy
$T_H<T$ and consequently it will accrete plasma rather than decaying with
plasma emission.  At a timescale $t\sim 1 {\rm TeV}^{-1}$ its will
acquire mass and reach the $M=M_0$ limit. Then it will not
be able to expand into the $w$ direction anymore, so beyond this point its  horizon area will be
of the size $A\simeq4\pi R^2z_c$, rather than $A=2\pi^2R^3$. Since the
relationship between mass and radius is unchanged, $R^2\simeq 8M/3\pi
M_5^3$, the area of the horizon and the entropy of the black hole will
be proportional to $M$. Then, unlike for AdS black holes, where $S\sim
M^{3/4}$, for RS1 black holes in the above mass range, $S\sim M$.  Then the
Hawking temperature remains constant $T_H\sim $0.2TeV during further
accretion.  These black holes could, in principle, grow until the
temperature of the outside world catches up with them.

The heuristic scenario outline above can be made more precise by solving
the Einstein equation in a perturbative fashion, in an asymptotic
expansion in
   powers of
$r^{-2}$. The procedure is similar to the one previously applied to
investigate corrections to 4 dimensional black hole solutions in brane
worlds~\cite{italian}. The details of this calculation will be presented
in a future publication.~\cite{ournext} The results of our study can be
summarized as follows: An ansatz for the
  metric tensor is taken as
\begin{eqnarray*}
ds^2&=&\left(\frac{z_c}{z_c-|w|}\right)^2(g_{tt}dt^2-g_{ww}dw^2-g_{rr}dr^2
\\
&-&2g_{rw}drdw-r^2d\Omega^2),
\end{eqnarray*}
where
\[
g_{ii}=1+\sum_{n=1}f_i^n(w)/r^{2n},
\]
where $i=t,w,r$
and
\[
g_{rw}=\sum_{n=1}f_{rz}^n/r^{1+2n}.
\]
The power of $r$ in the leading terms was chosen such that the
solution would go over smoothly into the Myers-Perry black hole solution
when the mass is sufficiently small.  The coefficient of
$r^2d\Omega^2$ was chosen to be unity.  This choice establishes the scale
of the radial coordinate $r$.  This choice is also consistent with
the Myers-Perry solution at $w<<z_c$.  Finally, in agreement with the
Myers-Perry solution we set $f_{w}^1=f_{rw}^0=0$.  Then we solve the
Einstein equation order by order of the asymptotic series\cite{footnote2}.
We also impose the junction conditions  for empty brane\cite{junction}. 
Then we obtain a unique solution in second order:
$f_{r}^1=-f_{t}^1=8M/3\pi M_5^3$. To make $f^2_i$
unique one has to fix the scale for the $w$ coordinate, as well. This can
be done by requiring that the leading contributions, $f_{w}^2$ and
$f_{rw}^1$  are exactly equal to the ones predicted by the Myers-Perry
solution.  Then one obtains
\[
f_t^2(w)=\frac{8M}{3\pi M_5^3}\left(-w^2+\frac{2w^3}{3z_c}\right).
\]
    Now the area of the horizon is given by
\begin{equation}
A=4\pi\int_0^{z_c-\lambda}dw\,
r^2(w)\,\theta(r(w))\sqrt{1+(dr/dw)^2},\label{area}
\end{equation}
where $r(w)$ is the radius of the horizon, which is the solution of the
equation
$g_{tt}=0$.  The result of this calculation is, as expected, that $A\sim
M^{3/2}$ for
$M<< M_{\rm 0}$, while $A\sim M +c + O(M^{-1})$ for $M>M_0$, where
$c<0$ is a calculable constant.  We call the region $M<M_0$ the low mass
region and
$M_0<M<M_{\rm crit}$ the intermediate region. $M_{\rm crit}$ will be
determined below.

Let us consider now the consequences of the dependence of
the entropy of the black hole on mass. Using (\ref{area}) we obtain the
following form for the entropy in the intermediate region ($M>>M_0$)
\begin{equation}
S\simeq \frac{8}{3}
Mz_c-\frac{3\pi}{20}(M_5z_c)^3+O(M^{-1}).
\label{entropy3}
\end{equation}
As the Hawking temperature is constant primordial black holes might
accrete matter until the universe cools to $T=T_H$. We will see later
that this will not really happen.  Still, assuming such an accretion is
possible it is interesting to see  at what mass value would the entropy
of the 5 dimensional black hole, (\ref{entropy3}), equal to the
entropy $S_4 = {4\pi M^2\over M_4^2}$ of a 4
dimensional black hole with the same mass. Equating these two expressions one obtains
\[
M_{\rm crit}\simeq \frac{2}{3\pi^2}M_4^2z_c\sim 10^{32}{\rm TeV}.
\]
It turns out that the Schwarzschild radius of such a four dimensional
black hole is $R=O({\rm TeV}^{-1})$ and its Hawking temperature is
$T_H=O($0.1TeV). This is a region where quantum effects are supposed to
enter in a world with a Planck mass of $M_5\simeq $1TeV.  Quantum
instability is supposed to transform the 4 dimensional black hole into
a 5 dimensional black hole as a Gregory-Laflamme type instability sets
in.   Though an $N$-black hole configuration has almost the
same entropy due to the linear dependence of the entropy on mass, the
negative constant term makes the decay into a single 5 dimensional black
hole slightly more favorable. We shall see that this does not happen.

The first warning sign about the validity of the above simple minded
scenario comes from considering the radii of these black holes.  While
the radius of the horizon of the four dimensional black hole is
$R\sim
$1TeV$^{-1}$, the decay product, the 5 dimensional black hole
has a macroscopic, $R\sim M_4/M_5^2\simeq 10^{16}{\rm
TeV}^{-1}\simeq$1mm radius of horizon.  It is physically difficult to
imagine such a transformation.  It would be aesthetically much more
pleasing  if the
$R=$1TeV$^{-1}$ four dimensional black hole would decay into black holes
of similar radius.  To see that this is what really happens we should
realize that  black holes of the intermediate region, $M_0<M<M_{\rm
crit}$ are not stable themselves. We can envision this in the following manner. As
soon as the size of black holes reaches the size of space in the 5th
dimension when their masses reach $M_0$,  they are forced to expand along the brane only. 
 As soon as
they are large enough along the brane to break up into two   more or
less  spherically symmetric 5 dimensional black holes, that barely fit the
fifth dimension, they will do so, thereby increasing their entropy.  In
fact, if we consider the relationship of entropies of $n$ 5 dimensional
spherical black holes and one flattened 5 dimensional black hole then we
obtain, in a rough approximation
\[
nS_5(M/n)=n\sqrt{2\pi}\left(\frac{4M}{3 n M_5}\right)^{3/2}= 8Mz_c/3
\]
Then we find that for every $n$ we find a mass $M_n$ for which this
equality is satisfied
\[
M_n=nM_5^3z_c^23/2\pi\sim nM_0.
\]
For $M>M_n$ a black hole of the intermediate region is unstable against
decaying into $n$ five dimensional black holes.

Then the fate of 4 dimensional black holes is also clear. As soon as they
reach the critical mass of $M_{\rm crit}$ where their radius is $R\sim
1{\rm TeV}^{-1}$  and Hawking temperature $T_H\simeq $1TeV they become
unstable due to quantum fluctuations and decay into a large number ($N$)
of 5 dimensional black holes,
\[
N\simeq M_{\rm crit}/M_0\simeq
\frac{16}{9\pi^{3}}\frac{M_4^2}{M_5^3z_c}\sim 10^{29}.
\]
The radius of the horizon of these black holes is $R\sim z_c$, very
similar to that of the decaying 4 dimensional black hole. Similarly, all
the black holes have a Hawking temperature of a fraction of 1TeV.

The moral of these considerations is that in RS1 there are two distinct
types of black holes that have {\em nothing to do with each other}:  4
dimensional ones that become unstable at a minimal mass, $M\sim
10^{32}$TeV, at which  the
size of their horizon is reduced to the quantum scale (the inverse of the
mass of KK modes) and 5 dimensional black holes that have maximal masses
of
$M\sim10^3$TeV. In between these two extremal masses there is no
stable black hole centered at the TeV brane.

Note that the instability we have studied
    differs substantially from that of Gragory and
Laflamme~\cite{gregory1, gregory2},  which transforms 4 dimensional black
holes into a five dimensional black hole of the same mass (not considering
    intermediate states). The reason for the difference is that  in
flat space, the heavier 5 dimensional black holes are, the
more stable they become. Contrast this behavior with what we have studied
in  RS1, where  5 dimensional black holes  of the
size $R\sim z_c$ are the most stable.

If after inflation the universe reheats to
$T\simeq$1TeV then one expects that a large number of 5 dimensional brane black
holes are produced by the collision of standard model particles. The
Hawking temperature of these black holes is lower than  the ambient
temperature and they are expected to rapidly accrete matter until they
reach $M\simeq 2M_0\sim 10^3$TeV, when they decay into two black
holes of mass $M_0$. This process then repeats itself until most of the
particles are transformed into black holes.  The black holes will
start to decay only after the outside temperature reaches their Hawking
temperature, in about $10^{-13}$s.

Finally let us briefly consider RS2 (single brane world) black holes. Here
the physical brane is at $y=0$
and the second brane is taken to infinity. In conformal coordinates z
   it is obvious
that for large $|z|$, far away from the brane,  the conformal factor
$\lambda^2/(\lambda+|z|)^2\simeq\lambda^2/z^2$ and the metric approaches
that of the AdS space {\em without a brane}. Therefore, far away from the
brane, black hole solutions of
the AdS type should exist. Conversely, when
$|z|<<\lambda$ then the space looks flat. Thus, close to the brane Myers
and Perry type black holes should exist.
We are interested in  black holes centered on the brane.

It was shown by Randall and Sundrum~\cite{rs2} that normal Newtonian
gravity prevails on the brane at distances $R>>\lambda$.  This constraint
puts an upper limit on $\lambda = M_4^2/M_5^3\le 1{\rm mm}\simeq
10^{16}{\rm TeV}^{-1}\sim 10^{32}M_4^{-1}$. This restricts the five
dimensional Planck scale to
be between $10^5$ TeV and the 4-d Planck scale of $10^{16}$ TeV. Black hole masses are bounded below by this five dimensional
Planck scale. For
$R>>\lambda$,
  black holes, which are effectively 3+1 dimensional, with a small extension
to the bulk,  should  exist\cite{hawking2}, with the
standard connection between the mass and the radius of the horizon:
$R=2M/M_4^2$. Such pan cake shaped black hole solutions have been found
using an expansion technique\cite{italian}. The entropy of
such a black hole is
\begin{equation}
S=\frac{4\pi M^2}{M_4^2}.
\label{entropy4}
\end{equation}
where $M_4=(G_4)^{-1/2}.$

The evaporation of such four dimensional RS2 black holes is similar to ADD.  
The size of the evaporating black
hole   approaches the AdS curvature length scale,  $\lambda$, at a
mass $M_{\rm crit}=\lambda M_4^2\simeq 10^{32}M_4$. It can be shown   that at that point it
 becomes entropically favorable for it to decay into an
almost flat 5 dimensional black hole centered on the brane. This
phenomenon could also be studied using an expansion method, similar to
the one applied to RS1.

It would be interesting to discuss the stability and the possible growth of
TeV-range brane black holes from the point of view of the AdS-CFT
correspondence.  The growth of these black holes by accretion in the early
universe and the possibility of creating a baryon asymmetry by such
accreting black holes will be discussed elsewhere.

The authors are indebted to P. Argyres, M.Bowick, F.P.Esposito, A. Shapere and J. Terning for
valuable discussions. The support of the U.S. Department of Energy under
Grant No. DE-FG02-84ER40153 is gratefully acknowledged.


\begin{thebibliography}{10}
\bibitem{note} Unlike in ADD in RS1 the universe could be heated to a
temperature of a TeV. See Ref.[4].
\bibitem{footnote2} For convenience we rearrange the series, such that it
is an expansion in powers of $r_{(5)}^{-2}=(w^2+r^2)^{-1}$, rather than
$r^{-2}$.
\end{thebibliography}
\end{document}